\documentclass[aps,prl,twocolumn,a4paper,showpacs,groupedaddress]{revtex4}

\usepackage{graphicx}
\usepackage{amsmath}
\usepackage{amssymb}
\usepackage{psfrag}

\newcommand{\beq}{\begin{equation}}
\newcommand{\eeq}{\end{equation}}
\newcommand{\beqa}{\begin{eqnarray}}
\newcommand{\eeqa}{\end{eqnarray}}
\newcommand{\cond}{\Phi_0 (\mathbf{r})}

\newcommand{\hsp}{\hat{H}_{\text{sp}}}
\newcommand{\nt}{\tilde{n}}
\newcommand{\mt}{\tilde{m}}
\newcommand{\bfr}{\mathbf{r}}
\newcommand{\bfrp}{\mathbf{r'}}
\newcommand{\rt}{(\mathbf{r},t)}
\newcommand{\rpt}{(\mathbf{r'},t)}
\newcommand{\rrt}{(\mathbf{r},\mathbf{r'},t)}
\newcommand{\rw}{(\mathbf{r},\omega)}

\newcommand{\intdr}{\int \!\! d \mathbf{r} \,}
\newcommand{\intdrp}{\int \!\! d \mathbf{r'} \,}
\newcommand{\col}[2]{\begin{pmatrix} #1 (\mathbf{r}) \\ #2 (\mathbf{r}) \end{pmatrix}}
\newcommand{\coldgPw}{\begin{pmatrix} \delta \Phi \rw \\ \delta \Phi^{*} (\mathbf{r},-\omega) \end{pmatrix}}
\newcommand{\bpm}{\begin{pmatrix}}
\newcommand{\epm}{\end{pmatrix}}

\begin{document}

\title{Quantitative test of thermal field theory for Bose-Einstein condensates}

\author{S. A. Morgan,$^1$ M. Rusch,$^2$ D. A. W. Hutchinson,$^3$ and K. Burnett$^2$}

\affiliation{$^1$Department of Physics and Astronomy, University College London, Gower Street, London WC1E 6BT, UK\\$^2$Clarendon Laboratory, Department of Physics, University of Oxford, Parks Road, Oxford OX1 3PU, UK\\$^3$Department of Physics, University of Otago, Dunedin, New Zealand }

\date{\today}

\begin{abstract}
We present numerical results from a full second order quantum field theory of Bose-Einstein condensates applied to the 1997 JILA experiment [D. S. Jin \textit{et al.}, Phys. Rev. Lett. {\bf 78}, 764 (1997)]. Good agreement is found for the energies and decay rates for both the lowest-energy $m=2$ and $m=0$ modes. The anomalous behaviour of the $m=0$ mode is due to experimental perturbation of the non-condensate. The theory includes the coupled dynamics of the condensate and thermal cloud, the anomalous pair average and all relevant finite size effects.
\end{abstract}

\maketitle

One of the most intriguing consequences of the experimental realization of Bose-Einstein condensation (BEC) was the prospect of quantitative tests of finite temperature quantum field theory (QFT). The pioneering measurements of condensate excitations at JILA provide the most stringent tests to date of such theories \cite{Jin96,Jin97}. Accurate calculations have proved difficult, however, because of the need to include the dynamic coupling of condensed and uncondensed atoms simultaneously with effects due to strong interactions and finite size. In this paper we describe the first direct comparison of a full second order QFT calculation with the JILA measurements. Our results show that accurate tests of QFT are possible if it is properly adapted to the finite, driven systems under consideration.

Measurements of excitations at low-temperature are in good agreement with predictions based on the Gross-Pitaevskii equation (GPE) and Bogoliubov quasiparticles
\cite{Jin96,Stringari96,Edwards96}. However, the finite-temperature JILA results \cite{Jin97} have proved much harder to explain. In this experiment the energies of the lowest-energy modes with axial angular momentum quantum numbers $m=2$ and $m=0$ were measured as a function of reduced temperature $t = T/T_c^0$, where $T$ is the absolute temperature and $T_c^0$ is the BEC critical temperature for an ideal gas. The $m=2$ mode was observed to shift downwards with $t$, while the $m=0$ mode underwent a sharp increase in energy at $t \sim 0.6$ towards the result expected in the non-interacting limit.

The temperature dependence of the excitations has been studied theoretically using the Popov approximation to the Hartree-Fock-Bogoliubov formalism, where the anomalous (pair) average of two condensate atoms is neglected. This gives good agreement with experiment for low temperatures but can not explain the results for $t>0.6$ \cite{Dodd98}. Good agreement for all $t$ for the $m=2$ mode was obtained using an extension of this approach which includes the anomalous average \cite{Hutchinson98}, and also within the dielectric formalism \cite{Reidl99}. However, both approaches were unable to explain the upward shift of the $m=0$ mode, and an analytical treatment of the problem also predicted downward shifts for both modes \cite{Giorgini00}. The importance of the relative phase of condensate and non-condensate fluctuations was emphasized by Bijlsma, Al Khawaja and Stoof (BKS) \cite{Bijlsma99}, who showed that the experimental results for $m = 0$ can be qualitatively explained by a shift from out-of-phase to in-phase oscillations at high temperature. Jackson and Zaremba (JZ) \cite{Jackson02} obtained good quantitative agreement for both modes using a GPE for the condensate coupled to a non-condensate modelled by a Boltzmann equation. However, this approach neglects the phonon character of low-energy excitations as well as the anomalous average and Beliaev processes. The anomalous average can be significant, especially near a Feshbach resonance \cite{Morgan00,Holland01} and Beliaev processes have been directly observed in a number of recent experiments \cite{Hodby01,Katz02,Bretin03}. It is therefore important to explain the JILA results using a theory which includes these effects.

In this paper we present numerical results for the excitations of a dilute gas BEC at finite temperature for the conditions of the 1997 JILA experiment \cite{Jin97}. We find good agreement with the experimental results for both the $m=2$ and $m=0$ modes, and in particular we are able to explain straightforwardly the anomalous behaviour of the $m=0$ mode. The results are based on a theoretical treatment recently developed by one of us (S.M), as an extension of an earlier second-order perturbative calculation \cite{Morgan03b,Morgan00}. The formalism adapts the linear response treatment of Giorgini \cite{Giorgini00} and closely models the experimental procedure where excitations are created by small modulations of the trap frequencies. The result is a gapless extension of
the Bogoliubov theory which includes the dynamic coupling between the condensate and non-condensate, all relevant Beliaev and Landau processes and the anomalous average. It is also consistent with the generalized Kohn theorem. The theory is valid in the collisionless limit of well-defined quasiparticles. For homogeneous systems at finite temperature this requires $(k_{\text{B}}T/n_{0}U_{0})(n_0a_s^{3})^{1/2} \ll 1 $, where $n_0$ is the condensate density, $a_s$ is the s-wave scattering length, $k_{\text{B}}$ is Boltzmann's constant and $U_0 = 4\pi\hbar^2a_s/m$ where $m$ is the atomic mass \cite{Giorgini00,Morgan00}. For the JILA experiment \cite{Jin97} this parameter does not exceed $0.03$ at the trap centre for the highest temperature we consider.

The theory starts from the generalized GPE for the condensate wavefunction $\Phi\rt$
\beqa
 i \hbar \frac{\partial\Phi }{\partial t} &=& \left [  \hsp + P\rt - \lambda(t) + N_0(t)U_0|\Phi|^2 \right ] \Phi \nonumber \\
&& + 2 U_0\nt\rt\Phi + U_0\mt\rt\Phi^* -f\rt.
\label{TDGGPE}
\eeqa
Here $\hsp = -\hbar^2\nabla^2/2m + V_{\text{trap}}(\bfr)$ is the static single-particle Hamiltonian, $P\rt$ is the time-dependent external perturbation and $\lambda(t)$ is a scalar. The non-condensate density $\nt\rt$, anomalous average $\mt\rt$ and $f\rt$ are constructed from time-dependent quasiparticle wavefunctions $u_i\rt$ and $v_i\rt$ by
\begin{align}
\nt \rt &= \sum_{i} |u_i\rt|^2 N_i +  |v_i\rt|^2 (N_i + 1), \label{nt_qp}\\
\mt \rt &= \sum_{i} u_i\rt v_i^*\rt (2N_i + 1),\label{mt_qp} \\
f\rt &= \sum_{i} c_i^*(t)N_i u_i\rt + c_i(t)(N_i+1)v_i^*\rt,\\
c_i(t) &= U_0 \intdr |\Phi|^2 \left [ \Phi^* u_i\rt + \Phi v_i\rt \right ].
\end{align}
The quasiparticle wavefunctions evolve according to
\begin{gather}
i \hbar \frac{\partial}{\partial t} \bpm u_{i} \\ v_{i} \epm = \bpm \hat{L} & \hat{M} \\ -\hat{M}^* & -\hat{L}^* \epm\bpm u_{i} \\ v_{i} \epm,\label{uvt}\\
\begin{align}
\hat{L}\rt &= \hsp + P\rt +N_0U_0 \left [ |\Phi|^2 + \hat{Q}|\Phi|^2\hat{Q} \right ],\\
\hat{M}\rt &= N_0 U_0\hat{Q}\Phi^2\hat{Q}^*,
\end{align}
\end{gather}
where the orthogonal projector $\hat{Q}\rrt = \delta(\bfr-\bfrp)-\Phi\rt\intdrp\Phi^*\rpt$ ensures orthogonality of the condensate and non-condensate.

The quasiparticle populations $\{ N_i \}$ are independent of time and given by the Bose-Einstein distribution $N_i = 1/(e^{\epsilon_i/k_{\text{B}}T}-1)$ where $\epsilon_i$ is the Bogoliubov energy (see below). Most quantities in the theory depend on temperature via these populations. The condensate population $N_0(t)$ is defined in terms of the fixed total number of particles $N$ by $N_0(t) = N-\intdr \nt\rt$. The zero-temperature part of the anomalous average $\mt\rt$ is ultra-violet divergent, but it can be renormalized straightforwardly \cite{Morgan00,Giorgini00}.

The above equations are obtained using the number-conserving approach of Castin and Dum, modified for finite temperature calculations \cite{Castin98,Morgan03b}. The terms $f\rt$ and $\hat{Q}$ are a feature of this approach and do not appear in symmetry-breaking theories. We find that they can give a significant contribution to the energy shifts.

In the static case, Eq.~(\ref{TDGGPE}) has a time-independent solution $\Phi \rt = \Phi(\bfr)$ which satisfies
\beqa
 \left [ \hsp -\lambda + N_0U_0 |\Phi(\bfr)|^2 + 2 U_0 \nt(\bfr) \right ]\Phi(\bfr) &&  \label{TIGGPE}\\
 + U_0\mt(\bfr)\Phi^*(\bfr) -f(\bfr) &=& 0,\nonumber
\eeqa
where $\lambda$ is the condensate eigenvalue, roughly equal to the chemical potential. If we set $\nt$, $\mt$ and $f$ to zero, we obtain the usual GPE, with wavefunction $\cond$ and energy $\lambda_0$. We solve Eq.~(\ref{TIGGPE}) by linearizing the change in energy and shape relative to this solution. Writing $\Phi \rightarrow \cond$ in Eq.~(\ref{uvt}), we obtain static quasiparticle wavefunctions $u_i\rt = u_i(\bfr)e^{-i\epsilon_i t/\hbar}$, $v_i\rt = v_i(\bfr)e^{-i\epsilon_i t/\hbar}$ and the Bogoliubov energies $\{\epsilon_i\}$. These solutions are used to construct $\nt(\bfr)$, $\mt(\bfr)$ and $f(\bfr)$ and also provide a convenient basis for the subsequent calculation.

The external perturbation $P\rt$ leads to all quantities developing a small time-dependent oscillation around their static values, $\Phi\rt = \Phi(\bfr) + \delta \Phi \rt$, $\nt \rt = \nt(\bfr) + \delta\nt\rt$, etc. Substituting these expressions into Eq.~(\ref{TDGGPE}) and linearizing, we obtain the equation of motion for the condensate fluctuation $\delta \Phi\rt$. This equation can be solved by combining it with its complex conjugate, Fourier transforming and expanding the fluctuation in the static quasiparticle basis
\beq
\coldgPw = \sum_{i} b_i(\omega) \col{u_i}{v_i}.
\label{qp_expansion}
\eeq
The expansion coefficients $b_i(\omega)$ are directly related to the condensate density fluctuations $\delta n_0 = \delta (N_0|\Phi|^2)$, which are measured experimentally.

Dynamics of the non-condensate can occur via two distinct mechanisms; either it is driven directly by the perturbation or indirectly via the condensate. If we neglect the first possibility and assume that only the single mode `p' is excited, then $b_p(\omega)$ is given by
\beq
b_p(\omega) = P_{p0}(\omega){\cal G}_p(\omega+i\gamma).
\label{expansion_soln}
\eeq
Here $P_{p0}(\omega)$ is the matrix element for the generation of the excitation from the condensate and $i\gamma$ is a small imaginary part in the frequency (discussed below). The resolvent ${\cal G}_p(\omega)$ is defined in terms of a frequency-dependent self-energy by
\begin{align}
{\cal G}_p(\omega) &= \frac{1}{\hbar \omega - \epsilon_p - \Sigma_p(\omega)},
\label{G}\\
\Sigma_p(\omega) &= \Delta E_{p}^{(S)} + \Delta E_{p}^{(D)}(\omega).
\label{Epw}
\end{align}
$\Sigma_p(\omega)$ contains two types of energy shifts, static $(S)$ and dynamic $(D)$, corresponding to the different roles of the thermal cloud. The static term $\Delta E_{p}^{(S)}$ comes from interactions between a condensate fluctuation and the static non-condensate mean-fields. The dynamic term $\Delta E_{p}^{(D)}(\omega)$ describes the driving of non-condensate fluctuations by the condensate and their subsequent back action, which leads to damping and energy shifts of condensate excitations. The inclusion of this contribution leads to a gapless excitation spectrum \cite{Giorgini00,Morgan00}.

However, the non-condensate can also be excited directly by the external perturbation, and can then generate condensate excitations. This process therefore changes the effective excitation matrix element $P_{p0}$ and can be included by replacing ${\cal G}_p(\omega+i\gamma)$ in Eq.~(\ref{expansion_soln}) with the modified resolvent ${\cal R}_p(\omega+i\gamma)$, defined by
\beq
{\cal R}_p(\omega) = \left [ 1 + \frac{\Delta P_{p0}^{(S)}(\omega) + \Delta P_{p0}^{(D)}(\omega)}{P_{p0}(\omega)} \right ]{\cal G}_p(\omega).
\label{R}
\eeq
The important extra term here is $\Delta P_{p0}^{(D)}(\omega)$ which describes the generation of non-condensate fluctuations by the perturbation and their subsequent coupling to the condensate. $\Delta P_{p0}^{(S)}(\omega)$ describes the effect of changes in the static condensate shape [$\Phi_0(\bfr) \rightarrow \Phi(\bfr)$].

The detailed definition of $\Delta E_{p}^{(D)}$ and $\Delta P_{p}^{(D)}$ is lengthy and is given elsewhere \cite{Morgan03b,dynamicstructure}. We note here that they are both calculated as a sum over many Landau and Beliaev processes which are resonant whenever an energy matching criterion is satisfied. The parameter $\gamma$ in Eq.~(\ref{expansion_soln}) is required to keep $\Delta E_{p}^{(D)}$ and $\Delta P_{p}^{(D)}$ finite at the resulting poles. Its inclusion can be formally justified from the finite experimental resolution and its value is of order the inverse of the experimental observation time. Our numerical results are essentially independent of this parameter for physically relevant values.

If $\Sigma_p$ and $\Delta P_{p}^{(D)}$ are roughly independent of frequency, the energy shift can simply be calculated from the poles of $\cal{G}$, i.e. the solutions to $E_p = \hbar \omega_p = \mbox{Real}\left [ \epsilon_p+\Sigma_p(\omega_p) \right ]$, while the decay rate is given by $\Gamma_p = -\mbox{Imag}\left [\Sigma_p(\omega_p) \right ]/\hbar$. This situation arises in homogeneous condensates where an excitation couples to a continuum of decay channels and the resolvents are Lorentzians. For a finite system, however, $\Sigma_p(\omega)$ depends on frequency, and neither ${\cal G}_p(\omega)$ nor ${\cal R}_p(\omega)$ are perfect Lorentzians. In this case, we extract energies and decay rates by fitting $b_p(\omega)$ to a complex Lorentzian plus a constant ($\gamma$ is subtracted from the resulting decay rate). This corresponds to the experimental procedure of fitting a decaying sinusoid to the condensate density fluctuations in the time domain. The frequency dependence of $P_{p0}(\omega)$ is included as a (known) weight function in the fit to ensure that only the experimentally relevant range of frequencies is included.

We present numerical results for the parameters of the JILA experiment \cite{Jin97}. We consider a condensate of $6000$ $^{87}$Rb atoms in an anisotropic harmonic trap with radial and axial trap frequencies of $\omega_r = 2\pi \times 129$Hz, $\omega_z = 2\pi \times 365$Hz. The scattering length is $a_s = 110$ Bohr. The condensate population is fixed for all the temperatures considered, which is consistent with the experimental results for $t<0.9$. Zero-temperature effects have been included using the appropriate ultra-violet renormalization. The external perturbation has the form $P({\bf r},t) \propto r^2 \cos(m_p \phi-\omega_d t)\Theta(t)\Theta(T_d-t)$ where $r$ and $\phi$ are the radial and azimuthal angle coordinates, $\omega_d \approx \epsilon_p/\hbar$ is the central drive frequency, $\Theta(\tau)$ is the unit step function and $T_d = 14$ms is the drive time. The parameter $\gamma$ is taken to be $\gamma = 0.036\hbar\omega_r$ \cite{Jin97}.

For a fixed $N_0$ we first solve the static GPE of Eq.~(\ref{TIGGPE}) with $\nt = \mt = f = 0$ to obtain $\cond$. We then calculate and store the quasiparticle basis functions $u_i(\bfr)$ and $v_i(\bfr)$ and unperturbed energies $\epsilon_i$ from the static limit of Eq.~(\ref{uvt}) for all states up to an energy cutoff $E_{\text{cut}} \sim 130\hbar\omega_r$. Using these we can construct all static and dynamic terms, defined by sums and integrals over various functions of the quasiparticles. The numerical calculation is difficult because of the need to deal simultaneously with the phonon character of low-energy states and significant single-particle effects. We therefore use an accurate Gaussian quadrature scheme together with a large value of $E_{\text{cut}}$ and a semi-classical approximation at high energy \cite{Giorgini00}. The final results are converged to within $5 \times10^{-3}\hbar\omega_r$. Further details are given in \cite{Morgan03b}. 

\begin{figure}
\psfrag{ylabel1a}[][]{$E/\hbar \omega_r$ ($m=0$)}
\psfrag{ylabel1b}[][]{$E/\hbar \omega_r$ ($m=2$)}
\includegraphics[width=\columnwidth]{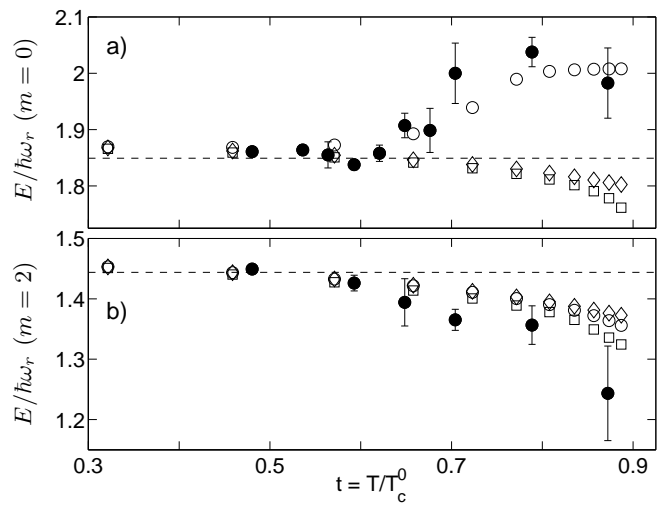}
\caption{Ab initio theoretical excitation energies $E$ (open symbols) compared with experiment (filled circles) for (a) $m_p = 0$ and (b) $m_p = 2$. Diamonds neglect direct thermal driving (${\cal G}_p$), open circles include it (${\cal R}_p$) and squares give $E_p$. Dashed line is the Bogoliubov energy $\epsilon_p$. Differences between diamonds and squares are due to non-Lorentzian structure in ${\cal G}_p$. There are no free parameters in the theoretical results.\label{energyfig}}
\end{figure}

Results for the $m=2$ and $m=0$ modes are compared to experiment in Fig~\ref{energyfig}.
As can be seen, the theory predicts a significant downwards shift for the $m=2$ mode. The agreement with experiment is reasonable if we consider the temperature error in the experiment (of order $5-10\%$) which is not shown. The downward curvature of the results is due to the scaling of the temperature axis from absolute to reduced temperature. For $k_{\text{B}}T \gg \lambda_0$ the shift is linear in $T$, as expected theoretically \cite{Giorgini00}.

\begin{figure}
\psfrag{ylabel2}[][]{$|{\cal R}_p(\omega)\times\hbar\omega_r|^2$}
\includegraphics[width=\columnwidth]{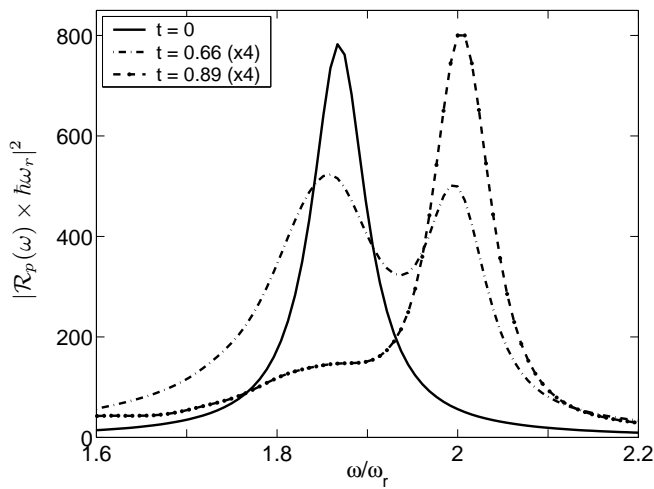}
\caption{Modulus squared of the dimensionless resolvent ${\cal R}_p(\omega) \times \hbar \omega_r$ as a function of frequency for $t=0$ (solid), $t=0.66$ (dot-dashed, $\times4$) and $t = 0.89$ (dashed, $\times4$).\label{resolventfig}}
\end{figure}

If we neglect thermal driving then similar behaviour is seen for the $m=0$ mode, as found in previous calculations \cite{Hutchinson98,Reidl99,Giorgini00}. Including this effect gives very different results, however, and the theory correctly reproduces the sharp upward shift in the excitation energy around $t = 0.6$. This is because an $r^2$ perturbation couples strongly to single-particle modes with frequency differences of $2\omega_r$ so the non-condensate response is peaked in this region. The effect on the condensate can be seen by plotting the modified resolvent $\cal{R}$ as a function of frequency and temperature as in Fig.~\ref{resolventfig}. The appearance of a growing peak at $\omega = 2\omega_r$ is due to direct driving of the non-condensate and is absent in an equivalent plot of $\cal{G}$.
In this case the perturbation mainly excites the non-condensate which then drives the condensate, rather than the reverse. This explanation of the experimental results is consistent with the conclusions of BKS and JZ \cite{Jackson02,Bijlsma99}. If the condensate drives the non-condensate the two oscillate out-of-phase. However, at high $t$ the non-condensate drives the condensate above its resonance frequency and hence the oscillations are in-phase. The out-of-phase branch should be observable using a perturbation localized around the condensate.

\begin{figure}
\includegraphics[width=\columnwidth]{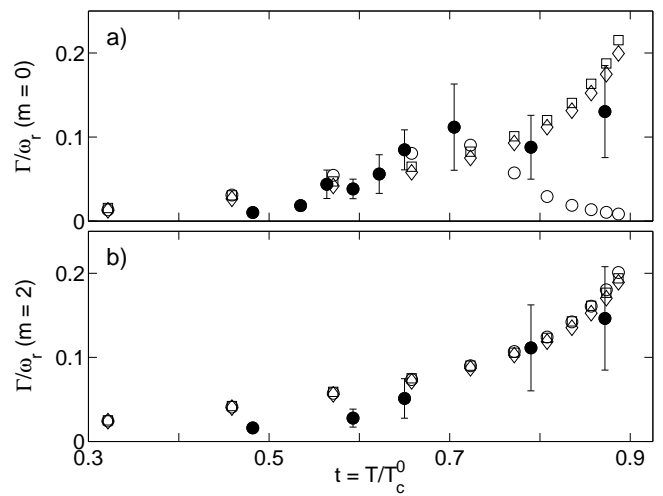}
\caption{Theoretical decay rates ($\Gamma$) compared with experiment for (a) $m_p = 0$ and (b) $m_p = 2$. Symbols as in Fig.~\ref{energyfig}.\label{decayfig}}
\end{figure}

Fig.~\ref{decayfig} shows the results for the damping rates. Overall the agreement with experiment is good, although the theory overestimates the damping rate at low temperatures. This was also seen by JZ \cite{Jackson02} and is possibly due to experimental difficulties in determining the temperature when the non-condensate fraction is small. For $m=0$, the damping rate is underestimated at high temperature if direct driving of the thermal cloud is included for reasons which are currently unclear.

In conclusion, we have presented numerical results from a gapless theory of condensate excitations which includes the anomalous average, Beliaev and Landau processes and the dynamic coupling of condensate and non-condensate fluctuations. Good agreement with the JILA experiment \cite{Jin97} is found for the energies and decay rates of the lowest-lying states with $m=2$ and $m=0$. This shows that a consistent perturbative approach is capable of explaining the experimental results, contrary to statements in the literature \cite{Reidl99,Jackson02}. The anomalous behaviour of the $m=0$ mode is the result of direct excitation of the non-condensate by the external perturbation.

S. M. and K. B. thank the Royal Society of London, the EPSRC and Trinity College, Oxford for financial support. D. A. W. H thanks the Marsden Fund of the Royal Society of New Zealand. S. M thanks M. J. Davis and S. A. Gardiner for many useful discussions.


\begin{thebibliography}{}
\bibitem{dynamicstructure} The structure of these terms is similar to Eq.~(80) of \cite{Morgan00}, but with different matrix elements.
\end{thebibliography}
\end{document}